\documentclass[oneside, a4paper, onecolumn, 11pt]{article}
\usepackage[left=2.5cm,top=2.5cm,bottom=2.5cm,right=2.5cm]{geometry}
\usepackage{amsmath}
\usepackage{graphicx}
\usepackage{fancyhdr}
\usepackage{calc}
\usepackage{amssymb}
\usepackage[super,sort&compress,comma]{natbib}
\usepackage{setspace}
\usepackage{amsfonts}
\usepackage{commath}
\usepackage[innercaption]{sidecap}
\usepackage[framemethod=TikZ]{mdframed}
\usepackage{hyperref}
\usepackage{parskip}
\usepackage{pifont}

\newcommand{\Rom}[1]{\expandafter\@slowromancap\romannumeral #1@}
\makeatother
{ \normalfont} 

\newcommand{\x}{\mathbf{x}}

\newcommand{\Basin}{\mathfrak{B}}

\newcommand{\Clique}{\mathcal{Q}}

\expandafter\def\expandafter\normalsize\expandafter{%
	\normalsize
	\setlength\abovedisplayskip{0pt}
	\setlength\belowdisplayskip{5pt}
	\setlength\abovedisplayshortskip{0pt}
	\setlength\belowdisplayshortskip{5pt}
}
\renewcommand{\baselinestretch}{1.1}
\definecolor{Gray}{gray}{0.75}

\newmdenv[backgroundcolor=Gray, leftmargin = 0pt, rightmargin = 0pt, linewidth = 0pt, roundcorner = 2 pt, innerleftmargin=5pt, innerrightmargin=5pt, innertopmargin=5pt, innerbottommargin=5pt]{Frame}

\begin{document}

\newcommand{\kk}{\langle k \rangle}
\newcommand{\kkk}{\langle k^2 \rangle}
\newcommand{\er}{Erd\H{o}s-R\'{e}nyi}
\newcommand{\red}{\color{red}\footnotesize}
\newcommand{\blue}[1]{{\color{blue} #1}}
\newcommand{\subfigimg}[3][,]{%

	\setbox1=\hbox{\includegraphics[#1]{#3}}
	\leavevmode\rlap{\usebox1}
	\rlap{\hspace*{30pt}\raisebox{\dimexpr\ht1-2\baselineskip}{#2}}
	\phantom{\usebox1}
}

\linespread{1.2}


\begin{center}
{\color{blue} \huge \textbf{Dense network motifs enhance dynamical stability}}

\vspace{2mm}
Bnaya Gross$^{1}$, Shlomo Havlin$^{1}$ \& Baruch Barzel$^{2,3}$
\end{center}

\small{
\begin{enumerate}
	\item
	\textit{Department of Physics, Bar-Ilan University, Ramat-Gan, Israel}
	\item
	\textit{Department of Mathematics, Bar-Ilan University, Ramat-Gan, Israel}
	\item
	\textit{Gonda Multidisciplinary Brain Research Center, Bar-Ilan University, Ramat-Gan, Israel}
\end{enumerate}

\vspace{4mm}

\textbf{Network motifs are the building blocks of complex networks \cite{milo2002network,kashtan2004efficient} and are significantly involved in the network dynamics such as information processing and local operations in the brain \cite{sporns2004motifs,sizemore2018cliques}, biological marks for drug targets \cite{zhang2016exploration,hwang2008identification}, identifying and predicting protein complexes in PPI networks \cite{liu2009complex,chen2014identifying,zhang2006identification,spirin2003protein,li2005interaction,palla2005uncovering,adamcsek2006cfinder,albert2004conserved}, as well as echo chambers in social networks \cite{baumann2020modeling,baldassarri2007dynamics,cinelli2021echo}. Here we show that dense motifs such as cliques have different stable states than the network itself. These stable states enhance the dynamical stability of the network and can even turn local stable states into global ones. Moreover, we show how cliques create polarization phenomena and global opinion changes.}

\pagenumbering{arabic}

Complex networks widely appear in nature and are governed by diverse and different dynamics and mechanisms \cite{boccaletti2006complex,barrat2008dynamical}. While many exotic phenomena such as chaotic behavior appear in many systems \cite{thompson1990nonlinear,aguirre2009modeling}, the most basic property of a complex system of any type is its stable states \cite{strogatz2018nonlinear}. Many nonlinear dynamics-driven networks often exhibit multiple stable states, often bi-stable, each surrounded by its \textit{basin of attraction}, a regime in the phase space that drags the system to the stable state once entering it. The boundaries of the basin of attraction describe how stable the state is, but finding them in the case of an $N$-dimensional system is an extremely challenging task. In this paper, we show how dense network motifs enhance the dynamical stability of a network and have significant advantages that can be utilized to control it in diverse scenarios.

\par

\vspace{5mm}
\textbf{\color{blue} \Large Network stability and perturbations}

Consider a complex network with $N$ nodes characterized by the adjacency matrix $A_{ij}$ and a degree distribution $P(k)$. $A_{ij} = 1$ imply that nodes $i$ and $j$ are connected and $0$ otherwise. The activity of node $i$, $x_i(t)$, accounts for the level of activity of the node in the network dynamics and can vary over time due to interactions with other nodes in the network or external interventions. The edge connecting nodes $i$ and $j$ indicate that the two nodes interact with each other such as two proteins in a PPI network or two species in an ecological network. The strength (weight) of the interaction, $\omega_{ij}$, follows a distribution $P(\omega)$ and indicates how significant the two nodes affect each other. Often the average weight $\omega = \sum_{\omega^{\prime} = - \infty}^{\infty} P(\omega^{\prime}) \omega^{\prime}$ is used for simplicity. The time evolution of the activity of node $i$ follow
\begin{equation}
    \dod{x_i}{t} = M_0(x_i) + \omega \sum_{j = 0}^N A_{ij} M_1(x_i) M_2(x_j) .
\label{eq:general}
\end{equation}

Here $M_0(x_i)$ account for the self-dynamics, i.e. how the node will behave if it was isolated, while the product $\omega M_1(x_i)M_2(x_j)$ describe the interaction between the node and each of his neighbours. \par
The fixed points of the system $\x_{\alpha} = (\x_{\alpha, 1},.....,\x_{\alpha,N-1})$ are obtained from Eq.~\eqref{eq:general} by setting the derivatives to zero. Each of the fixed points is surrounded by its basin of attraction

\begin{equation}
\Basin_{\alpha} = \Big\{ \x(t = 0) \,\, \Big| \,\, \x(t \rightarrow \infty) = \x_{\alpha} \Big\},
\label{Basin}
\end{equation}

a regime in the phase space that directs the system, if initiate within it, to $\x_{\alpha}$. \par

If the system has more than a single stable state (often bi-stable) one of them (usually the one with high activity) is preferred. In the case of bi-stable systems we denote the lower stable state as $\x_0$ and the higher as $\x_1$ with the basin of attractions  $\Basin_0$ and $\Basin_1$ respectively. If a system is initiate in $\Basin_0$ it will move toward $\x_0$ and if it is initiate in $\Basin_1$ it will move toward $\x_1$. Thus, the larger the basin of attraction is, the more stable is the state. \par

As a benchmark for studying network stability we consider three different dynamics characterizing three classes of dynamics. The first is Cellular dynamics described using Michaelis–Menten kinetics \cite{johnson2011original} which is a representative of all dynamics with $\x_1>0$ and $\x_0 = 0$ (Fig.~\ref{fig:dynamical stability}\textbf{a})
\begin{equation}
\dod{x_i}{t} = Bx_i^a + \omega \sum_{j = 0}^N A_{ij} \frac{x_j^h}{1+x_j^h}.
\label{eq:general_cellular}
\end{equation}
The second is Neuronal dynamics using the Wilson–Cowan model \cite{wilson1972excitatory,wilson1973mathematical,laurence2019spectral} representing classes with $\x_1>0$ and $\x_0 > 0$ (Fig.~\ref{fig:dynamical stability}\textbf{b})
\begin{equation}
\dod{x_i}{t} = -x_i + \omega \sum_{j = 0}^N A_{ij} \frac{1}{1 + e^{\mu -\delta x_i}}.
\label{eq:neuronal}
\end{equation}

and finally Social dynamics \cite{baumann2020modeling} which represents classes with $\x_1>0$ and $\x_0 < 0$ (Fig.~\ref{fig:dynamical stability}\textbf{c})
 \begin{equation}
 \dod{x_i}{t} = B -\gamma x_i + \omega \sum_{j = 0}^N A_{ij} \tanh{[\alpha(x_j - \mu)]}.
 \label{eq:social}
 \end{equation}
 
Since $\x_1$ is the preferred state we are interested in how stable it is i.e. if the system is perturbed to $\x_1 + \vec{\Delta}$ will it stay in $\Basin_1$? This question is not simple since the system is $N$-dimensional and it is very hard to find the boundaries of $\Basin_1$ especially for real-world dynamics which usually doesn't characterize by a Lyapunov function \cite{chiang1989stability}. One approach to counter this problem is to reduce the system to 1$D$ by considering only the average activity $\x$. In such cases Eq.~\eqref{eq:general} take the form \cite{gao2016universal}
\begin{equation}
\dod{\x}{t} = f(\x) = M_0(\x) + \beta M_1(\x) M_2(\x) .
\label{eq:general_MF}
\end{equation}
where $\beta$ is the average weighed degree of the network. \par

As can be seen in Fig.~\ref{fig:dynamical stability}\textbf{a-c}, the number of steady states depend on $\beta$. Here we will focus on the regime where the network has two stable states. The stable states $\x_0$ and $\x_1$ are obtaining by setting $f(\x) = 0$ and demanding $\dod{f}{\x} < 0$ (Fig.~\ref{fig:dynamical stability}\textbf{d-f}). The stability of states can be described using the dynamical potential \cite{ma2021universality}
\begin{equation}
	U(\x) = - \int_{0}^{\x} f(\x') d\x', 
	\label{eq: U_general}
\end{equation}
where the stable states corresponds to local minima. The potential gap $\Delta U$ between a stable state and the near maxima describe how stable the state is. As shown in Fig.~\ref{fig:dynamical stability}\textbf{g-i}, $\x_1$ gets more stable for denser networks.

\vspace{5mm}
\textbf{\color{blue} \Large Clique dynamics}

Dense network motifs exist in many real-world networks (Fig.~\ref{fig:Networks_motifs}\textbf{a-h}). Their distinct structure is usually denser compared to the network itself and with higher activity compared to the average (Fig.~\ref{fig:Networks_motifs}\textbf{i-m}). Thus, their dynamics is different then the dynamics of the average (Eq.~\eqref{eq:general_MF}) with different effects on the network stability. To describe the dynamics of motifs we will use a \textit{cliques} as a representative structure but our results are valid for all dense motifs.
\par
A clique $\Clique$ is a subset of nodes of size $q = |\Clique|$ where every pair of nodes in $\Clique$ are connected (Fig.~\ref{fig:Networks_motifs}\textbf{a-c}). We study effect of clique on the network stability by describing the dynamics of the clique activity $\x_q$ which follow (see SI)

\begin{equation}
      \dod{\x_q}{t} = f_q(\x_q) =  M_0(\x_q) +  M_1(\x_q) \big [ \beta_{q,in} M_2(\x_q) + \beta_{q,out} M_2(\x) \big ] ,
      \label{eq:general_clique_xq}
\end{equation}

where  $\beta_{q,in} = \omega(q-1)$ and $\beta_{q,out} = \omega(\kappa_q - q+1)$ are the average weight within and outside the clique of a single node and $\kappa_q$ is the average degree of nodes in the clique (see SI).

\par
The fixed points of the clique are obtained by setting the derivative in Eq.~\eqref{eq:general_clique_xq} to zero. The clique's high stable state $\x_{q,1}$ is obtained by setting $\x = \x_1$ and the low stable state $\x_{q,0}$ by setting $\x = \x_0$ (Fig.~\ref{fig:cellular_stability}\textbf{a}). The high stable state $\x_{q,1}$ increases with the clique size (Fig.~\ref{fig:cellular_stability}\textbf{b}) and the stability of the clique can be describe using the potential (Fig.~\ref{fig:cellular_stability}\textbf{c})

\begin{equation}
	U_q(\x_q) = - \int_{0}^{\x_q} f_q(\x_q') d\x_q',
	\label{eq:U}
\end{equation}
where we set the activity of the network outside the clique to $\x_0$ to isolate the dynamics of the clique from the support of the network. \par
Interestingly, as the clique size increases, the potential gap $\Delta U_q$ increases as well (Fig.~\ref{fig:cellular_stability}\textbf{d}) making $\x_{q,1}$ more stable. The direct implication of $U_q$ is that as long as the clique is within the potential well, even if the entire system is perturbed to $\x_0$, the clique will rise to $\x_{q,1}$ and drag the network back to $\x_1$ (Fig.~\ref{fig:cellular_stability}\textbf{e-f}). Thus, the effect of the clique on the network stability can be describe by $\Delta_c$, the largest clique perturbation $\x_q = \x_{q,1} - \Delta_c$ that will still bring the network back to $\x_1$. \par

To find $\Delta_c$ we need to find the corresponding clique activity $\x_q$ fitting the maxima of $U_q$. This can be found from the condition

\begin{equation}
\begin{cases}
R(\x_{q,1} - \Delta_c) = \beta_{q,out}M_2(\x_0),
\\[10pt]
\dod{f(\x_q)}{\x_q} \bigg |_{\x_q = \x_{q,1} - \Delta_c} > 0,
\end{cases}
\label{eq:general_R_deltac}
\end{equation}
where

\begin{equation}
R(x) = -\beta_{q,in} M_2(x) - \frac{M_0(x)}{M_1(x)}.
\label{eq:general_R}
\end{equation}

%
%
%
%

\textbf{Example I: Cellular dynamics.--} To illustrates this phenomena let us consider the cellular dynamics in Eq.~\eqref{eq:general_cellular}. In this case the clique dynamics described in Eq.~\eqref{eq:general_clique_xq} takes the form

\begin{equation}
\dod{\x_q}{t} = -B\x_q^a + \beta_{q,in} \frac{\x_q^h}{1 + \x_q^h} + \beta_{q,out} \frac{\x^h}{1 + \x^h}.
\label{eq:cellular_clique_xq}
\end{equation}
 \par
Thus, $\Delta_c$ can be obtained from Eq.~\eqref{eq:general_R_deltac} as 

\begin{equation}
- \beta_{q,in}\frac{(\x_{q,1} - \Delta_c)^h}{1 + (\x_{q,1} - \Delta_c)^h} +B(\x_{q,1} - \Delta_c)^a  = 0.
\label{eq:cellular_deltac}
\end{equation}

where Eq.~\eqref{eq:general_R} takes the form
\begin{equation}
R(x) = -\beta_{q,in} \frac{x^h}{1 + x^h} + Bx^a.
\label{eq:R_cellular}
\end{equation}

In the limit of large dense cliques $\beta_{q,in} \gg 1$, both $\Delta_c$ and $\x_{q,1}$ follow asymptotic behaviour 

\begin{equation}
\x_{q,1}, \Delta_c \sim \beta_{q,in}^{\frac{1}{a-h}}.
\label{eq:cellular_deltac_scaling}
\end{equation}
as shown in Fig.~\ref{fig:cellular_stability}\textbf{b,g}.
\par
Interestingly, larger clique size does not necessarily guarantee higher $\Delta_c$ and therefore higher stability as shown in Fig.~\ref{fig:cellular_stability}\textbf{g} for Human and Yeast PPI networks. The reason is that being close to many cliques increases the clique stability even if the clique itself is not so large.
\par
Since a node can be part of many cliques we need a measurement to estimate its contribution to the network stability. To do so we will define the stability as the sum of maximal perturbation of all cliques that a node is part of
\begin{equation}
	S_i = \sum_{\Clique, i \in \Clique} \Delta_c(\Clique) .
	\label{eq:S}
\end{equation}
Fig.~\ref{fig:cellular_stability}\textbf{h-k} shows the stability of nodes in Human and Yeast PPI networks. As can be seen the distribution of nodes stability $P(S)$ follows a power-law behaviour indicating that a few nodes are the most important nodes that contribute to the networks stability.

\vspace{5mm}
\textbf{\color{blue} \Large Global stability}

\par
In cases where $\x_0 > 0$, a critical clique size $q_c$ exists which above it the low stable state  $\x_{q,0}$ disappear and $\x_{q,1}$ becomes globally stable (Fig.~\ref{fig:Neuronal_stability}). In such cases the clique spontaneously rises to $\x_{q,1}$ for any initial condition and drag the network to $\x_1$, making $\x_1$ globally stable as well. Thus, while $\Delta_c$ describe the \textit{local} stability of $\x_1$, $q_c$ describe its \textit{global} stability. The clique critical size can be found by demanding $\x_q$ to be half stable

\begin{equation}
\begin{cases}
R(\x_{q_c,1} - \Delta_c) = \beta_{q, out} M_2(\x_0),
\\[10pt]
\dod{f(\x_q)}{\x_q} \bigg |_{\x_q = \x_{q_c,1}- \Delta_c} = 0,
\end{cases}
\label{eq:general_R_deltac_qc}
\end{equation}
where $\x_{q_c,1}$ is the clique fixed point for $q_c$.

%
\par

This phenomena can happen only when $\x_0 > 0$ due to the supportive nature of the clique. If $\x_0 = 0$ the nodes in the clique will not support each other without activation and the system will stay at $\x_0$ as shown for Cellular dynamics (Fig.~\ref{fig:cellular_stability}). 
\par
\textbf{Example II: Neuronal dynamics.--} To demonstrate this phenomena let us consider the neuronal dynamics in Eq.~\eqref{eq:neuronal}. The clique can be obtained from Eq.~\eqref{eq:general_clique_xq} as

\begin{equation}
\dod{\x_q}{t} = -\x_q + \frac{\beta_{q,in}}{1 + e^{\mu -\delta \x_q}} + \frac{\beta_{q, out}}{1 + e^{\mu -\delta \x}}.
\label{eq:neuronal_clique_xq}
\end{equation}
The clique fixed points are obtained by setting Eq.~\eqref{eq:neuronal_clique_xq} to zero (Fig.~\ref{fig:Neuronal_stability}\textbf{a}). Since the clique is denser compare to the network itself its stable states are higher  ($\x_{q,0} > \x_0$ and $\x_{q,1} > \x_1$) as shown in Fig.~\ref{fig:Neuronal_stability}\textbf{b}. The stability of the clique can be described by the potential $U_q$ (Fig.~\ref{fig:Neuronal_stability} \textbf{c}). The potential gap $\Delta U_q$ increases with the clique size and making the system more stable. Interestingly, above a critical clique size $q_c$ the low stable state $\x_{q,0}$ disappear making $\x_1$ globally stable (Fig.~\ref{fig:Neuronal_stability} \textbf{d}).
\par
The clique critical size $q_c$ can be obtained from Eq.~\eqref{eq:general_R_deltac_qc} as the solution of the coupled equations

\begin{equation}
\begin{cases}
-\frac{\beta_{q_c,in}}{1 + e^{\mu -\delta (\x_{q_c,1} - \Delta_c)}} + \x_{q_c} - \Delta_c= \frac{\beta_{q_c, out}}{1 + e^{\mu -\delta \x_0}},
\\[15pt]
-\beta_{q,in} \delta e^{\mu - \delta (\x_{q_c,1} - \Delta_c)} = (1 + e^{\mu - \delta (\x_{q_c,1} - \Delta_c)})^2.
\end{cases}
\label{eq:neuronal_qc}
\end{equation}

where

\begin{equation}
R(x) = -\frac{\beta_{q,in}}{1 + e^{\mu -\delta x}} + x.
\label{eq:R_neuronal}
\end{equation}

\par
To demonstrate how forming a clique can spontaneously raise the system from $\x_0$ to $\x_1$ we initiate a system in $\x_0$ and created a clique of size $q = 3$. Then, we started increasing the clique size by adding more nodes to the clique. As shown in Fig.~\ref{fig:Neuronal_stability}\textbf{e}, as the clique size increases, its low stable state $\x_{q,0}$ increases as well but as long as the size of the new clique is below $q_c$ the network will stay at $\x_0$. However, once sufficiently large clique has been formed the system will spontaneously move to $\x_1$ making it globally stable.

\par
In cases where $q < q_c$ the cliques still affecting the local stability of $\x_1$ which described by $\Delta_c$. As shown in Fig.~\ref{fig:Neuronal_stability}\textbf{f} $\Delta_c$ increases as the cliques getting larger for $q < q_c$ making $\x_1$ more locally stable.

\vspace{5mm}
\textbf{\color{blue} \Large Polarization}

In cases where $\x_1 > 0$ and $\x_0 < 0$ such as in social dynamics (Eq.~\eqref{eq:social}), an interesting polarization phenomenon may happen where the activity in the network is spread among the stable states \cite{baumann2020modeling,baldassarri2007dynamics}. Here we show how cliques significantly affect this phenomenon and even make it disappear. 
\par
\textbf{Example III: Social dynamics.--} To demonstrate this phenomena let us consider the social dynamics. Cliques are widely common in social networks (Fig.~\ref{fig:Networks_motifs}\textbf{c}) and function as echo chambers where clique member exchange opinion intensively \cite{cinelli2021echo}. The clique dynamics of Eq.~\eqref{eq:social} can be obtained from Eq.~\eqref{eq:general_clique_xq} as

\begin{equation}
\dod{\x_q}{t} = B-\gamma \x_q + \beta_{q,in}\tanh[\alpha(\x_q - \mu)] + \beta_{q, out}\tanh[\alpha(\x - \mu)] .
\label{eq:social_clique_xq}
\end{equation}

The fixed points of the clique are obtained by setting Eq.~\eqref{eq:social_clique_xq} to zero. Interestingly, the clique stable states are more extreme than the stable state of the networks itself i.e. $|\x_{q,1}| > |\x_1|$ and $|\x_{q,0}| > |\x_0|$ as shown in Fig.~\ref{fig:Social_stability}\textbf{a-d}. The stability of the $\x_1$ can be studied using the potential $U_q(\x_q)$ (Fig.~\ref{fig:Social_stability}\textbf{e-f}) obtained from Eq.~\eqref{eq:U} where the potential gap $\Delta U_q$ increases with the clique size. To find $\Delta_c$ we can use Eq.~\eqref{eq:general_R_deltac} as

\begin{equation}
-\beta_{q,in}\tanh[\alpha(\x_{q,1} - \Delta_c - \mu)] - B + \gamma (\x_{q,1} - \Delta_c)  = \beta_{q, out}\tanh[\alpha(\x_0 - \mu)],
\label{eq:social_deltac}
\end{equation}

where Eq.~\eqref{eq:general_R} takes the form

\begin{equation}
R(x) = -\beta_{q,in}\tanh[\alpha(x - \mu)] - B + \gamma x .
\label{eq:R_social}
\end{equation}

\par

Here it is important to distinguish between biased ($B \neq 0$) and unbiased ($B = 0$) social dynamics. In the unbiased case where the activity of both states is equally strong $|\x_0| = |\x_1|$ the clique function as echo chamber and above $\x_{q,1} - \Delta_c$ will rise to $\x_{q,1}$, however, the network itself will stay at $\x_0$ (Fig.~\ref{fig:Social_stability}\textbf{g}). The reason for this is that the neighbours of the cliques are being pulled toward $\x_0$ by their neighbours stronger than they are being pulled by the clique. This create a polarization state between the clique and the rest of the network and the stability of $\x_1$ is no enhanced by the clique. This can explain how radical groups can sustain a radical opinion whilst the rest of the network is in the complete opposite opinion \cite{isenberg1986group}. \par

However, in the biased case $|\x_0| < |\x_1|$ one opinion is significantly more influencing then the other. Thus, the clique rise to $|\x_1|$ and pull its neighbours up since it over comes its neighbours due to its high influence (Fig.~\ref{fig:Social_stability}\textbf{h}). In this case cliques do enhance the stability of $\x_1$ and show that an echo chamber can alter the entire opinion of a social network if one of the state is significantly more influential then the other. This might explain how new trends and fashions can rapidly take over a population.
 
\par
\vspace{3mm}
\textbf{\color{blue} \Large Summary}

Cliques are common, dense structures that appear in many real-world systems. Their distinct structure makes their activity different compared to the entire network. This difference results in novel phenomena that depend on the network dynamics. On the basic level, cliques enhance the dynamical stability of high activity states, as shown for cellular dynamics, and can be used to efficiently revive failed networks \cite{sanhedrai2022reviving}. In other cases, when cliques of sufficient size are formed, local stable states can become globally stable and spontaneously attract the system, as shown for neuronal dynamics. Also, cliques can form polarization phenomena due to the echo chamber effect and have a different stable state compared to the entire network. Furthermore, if these cliques are extremely influential, it can also result in a global opinion change, attracting the entire system to their opinion. These properties make cliques important for external intervention, such as drag targets, where nodes that are part of many cliques are excellent candidates for these purposes.

\vspace{10mm}

{\color{blue} \textbf{Data availability}}.\
Empirical data required for constructing the real-world networks (Brain, Yeast PPI, Human PPI, Social) will be uploaded to a freely accessible repository upon publication.

{\color{blue} \textbf{Code availability}}.\
All codes to reproduce, examine and improve our proposed analysis will be made freely available online upon publication.

{\color{blue} \textbf{Acknowledgments}}.\
We thank the Israel Science Foundation, the Binational Israel-China Science Foundation Grant No.\ 3132/19, ONR, NSF-BSF Grant No.\ 2019740, the EU H2020 project RISE (Project No. 821115), the EU H2020 DIT4TRAM, and DTRA Grant No.\ HDTRA-1-19-1-0016 for financial support. B.G. acknowledges the support of the Mordecai and Monique Katz Graduate Fellowship Program.

{\color{blue} \textbf{Competing interests}}.\
The authors declare no competing interests. 

\clearpage

\renewcommand{\baselinestretch}{0.9}
\footnotesize
\bibliographystyle{unsrt}
\bibliography{bibliography}

\clearpage

\begin{figure}
\centering
\includegraphics[width=0.95\textwidth]{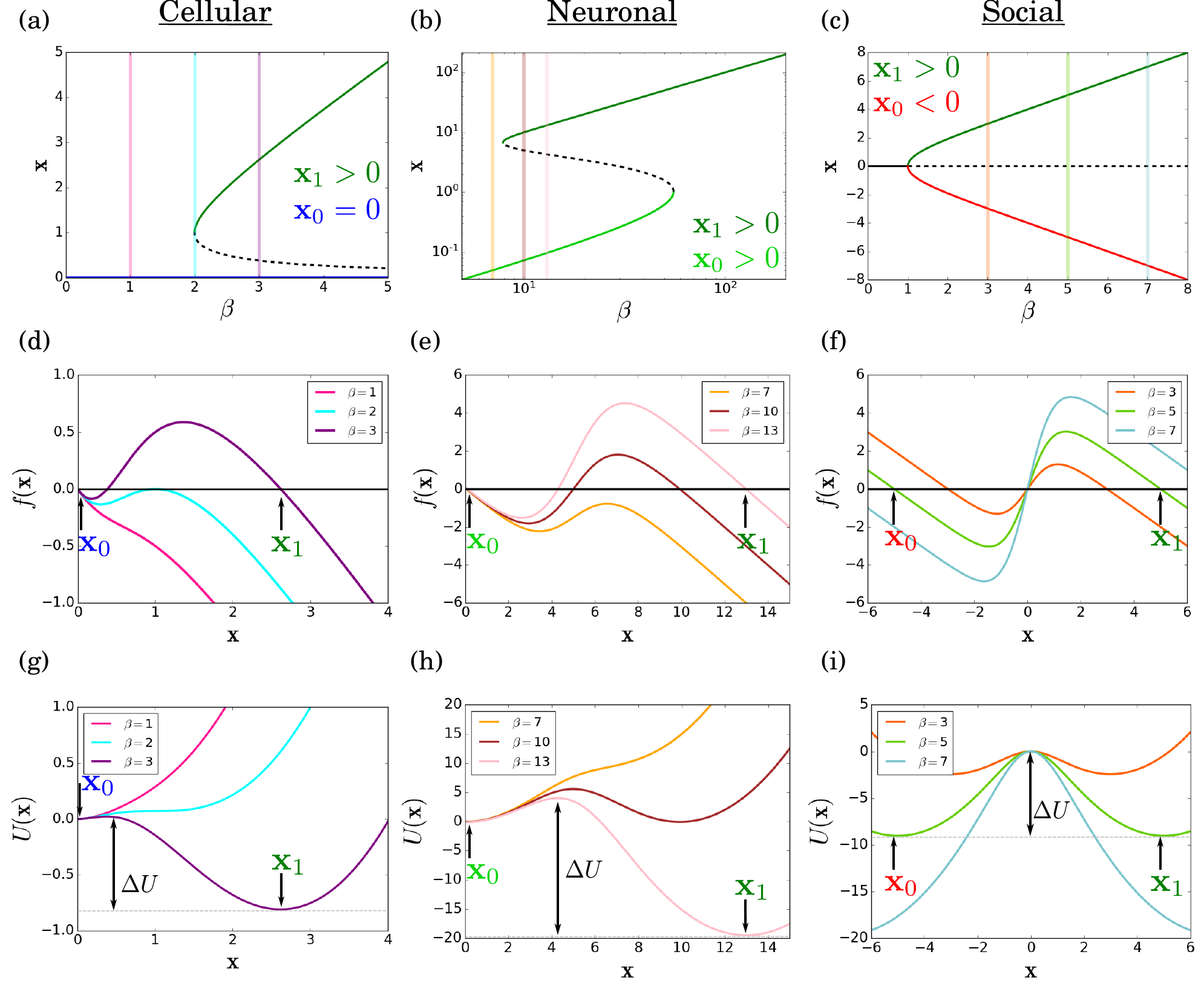}
\caption{\footnotesize \textbf{Dynamical stability.} A bi-stable system has a low stable state $\x_0$ and a preferred high stable state $\x_1$. Three examples are considered here for different cases of bi-stability: \textbf{(a)} Cellular dynamics with $\x_0 = 0$ and $\x_1 > 1$, \textbf{(b)} Neuronal dynamics with $\x_0 > 0$ and $\x_1 > 1$ and \textbf{(c)} Social dynamics with $\x_0 < 0$ and $\x_1 > 1$. \textbf{(d-f)} The stable states of each system are obtained by setting $f(\x)$ in Eq.~\eqref{eq:general_MF} to zero and depends on the average weighted degree of the network $\beta$. \textbf{(g-i)} The stability of the states can be described using the dynamical potential $U(\x)$ (Eq.~\eqref{eq: U_general}), where the potential gap $\Delta U$ between the stable state and the nearest maxima describes its stability.}

\label{fig:dynamical stability}	
\end{figure}

\clearpage

\begin{figure}
\centering
\vspace{-15mm}
\includegraphics[width=1\textwidth]{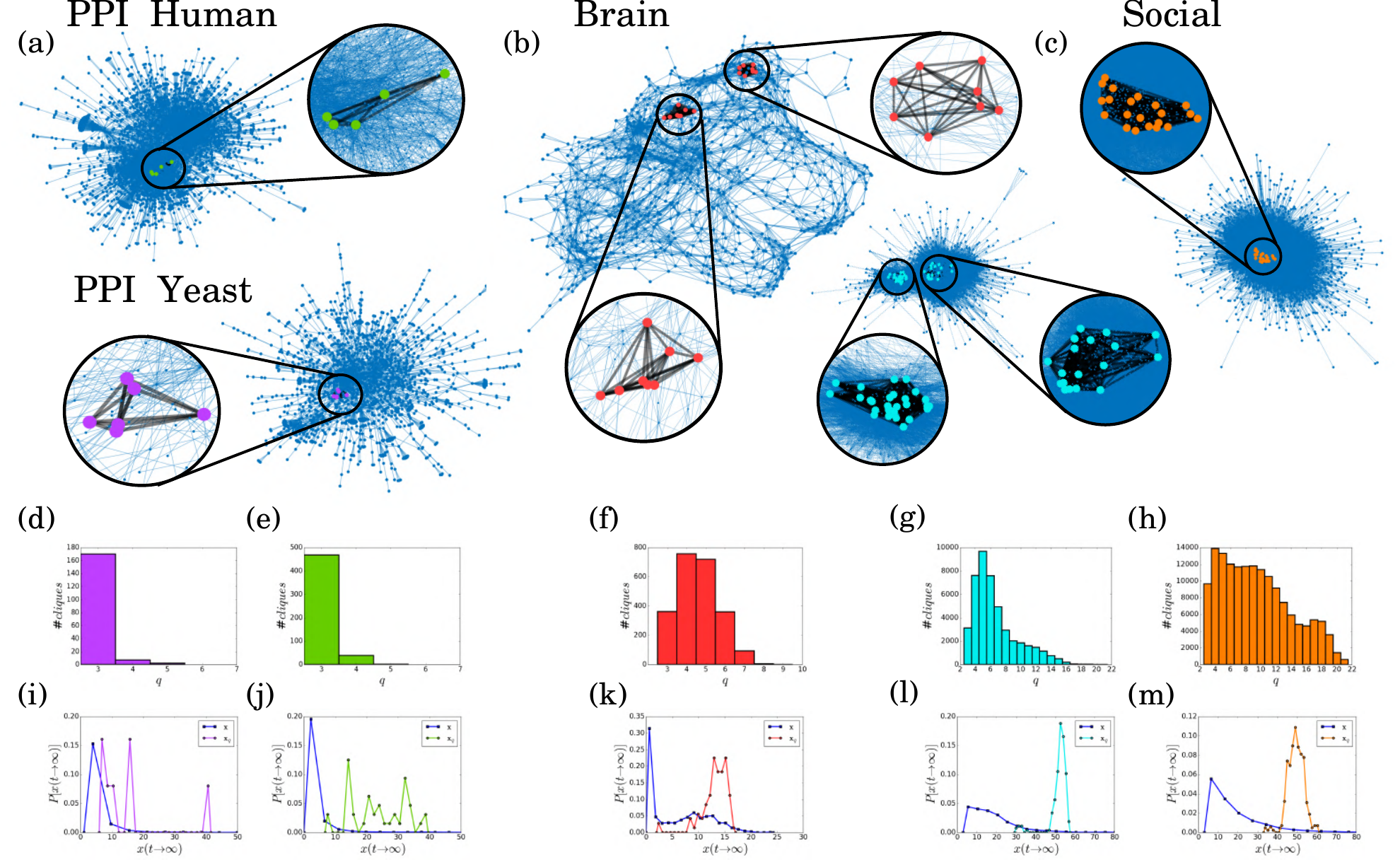}
\vspace{-5mm}
\caption{\footnotesize \textbf{Cliques in real-world networks. } Cliques are extreme examples of dense sub-structures appearing in many real-world networks governed by different dynamics, including \textbf{(a)} the PPI networks of humans and yeast, \textbf{(b)} the human brain, and  \textbf{(c)} social networks. \textbf{(d-h)} Cliques of different sizes exist in all of these systems, with different distributions. \textbf{(i-m)} The dense structure of cliques compared to the entire network results in higher activity of these dense structures ($\x_q > \x$) that enhance the dynamical stability of the network.}	
\label{fig:Networks_motifs}
\end{figure}

\clearpage

\begin{figure}
	\centering
	\includegraphics[width=0.95\textwidth]{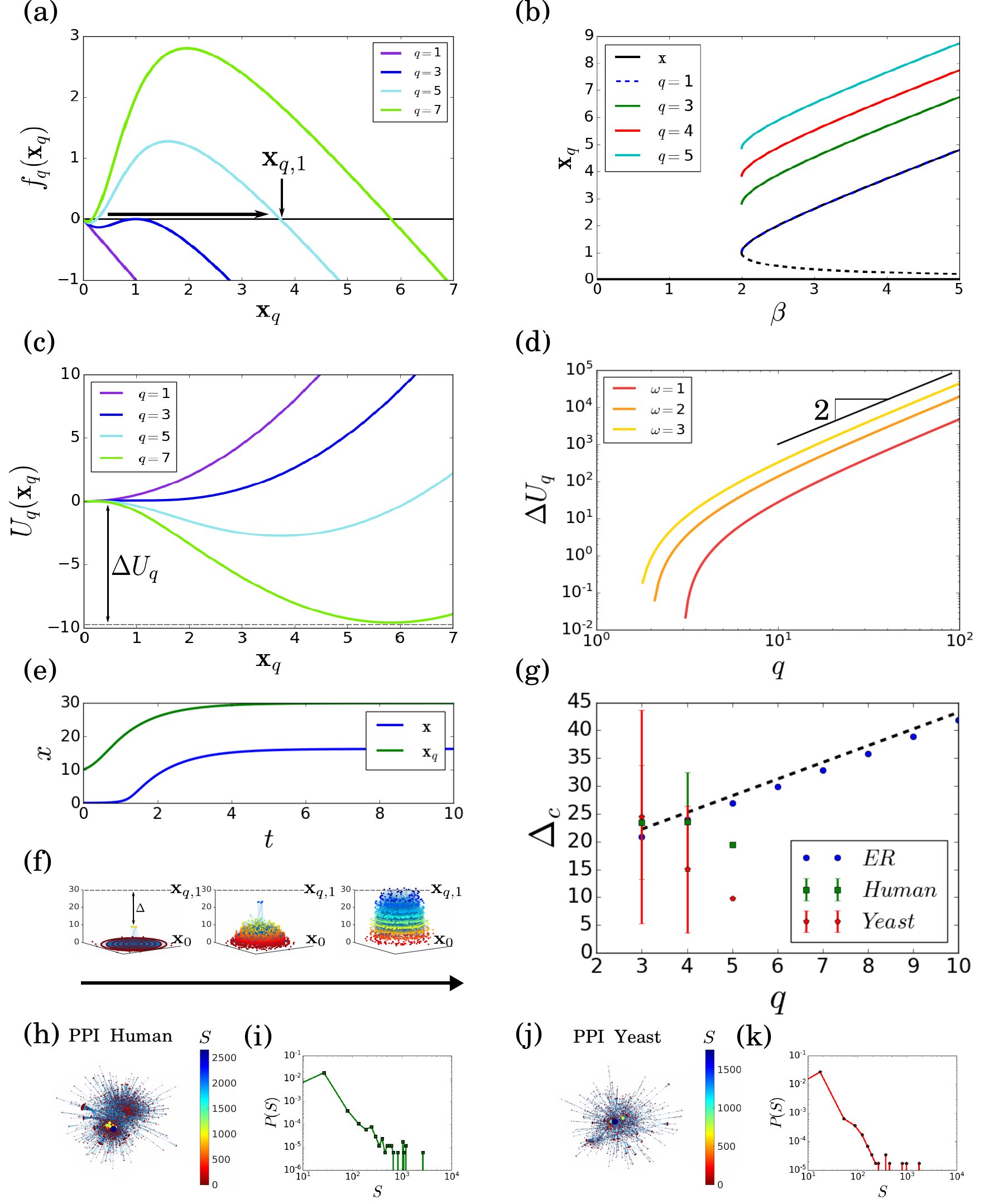}
	\caption{\footnotesize \textbf{Cellular dynamics stability. (a)} A new stable state $\x_{q,1}$ appears for large enough cliques, which is obtained by setting $f_q(\x_q)$ to zero in Eq.~\eqref{eq:cellular_clique_xq}. \textbf{(b)} The activity of the stable state $\x_{q,1}$ increases with the size of the clique and due to its dense structure.  \textbf{(c)} The stability of the new stable state can be measured using the clique potential gap $\Delta U_q$ to the nearest maxima, which increases with the clique size. \textbf{(d)} For large enough cliques, the scaling of the potential gap and the clique size follow $\Delta U \sim q^2$. \textbf{(e-f)} When the system is initiated at $\x_0$ and the clique activity remain in the basin of attraction of  $\x_{q,1}$ with activity higher than $\x_{q,1} - \Delta_c$, the clique's activity increases toward $\x_{q,1}$ while dragging the entire system to $\x_1$.  \textbf{(g)} The activity gap $\Delta_c$ increases with the clique size. However, in real-world cases this role does not necessarily apply since clique activity is influenced by its surroundings. Thus, isolated cliques will have lower activity than non-isolated cliques, even though their size is larger. \textbf{(h-k)} The contribution of a node to the stability of the network, $S_i$, depends on the number of cliques it is part of and is calculated from Eq.~\eqref{eq:S}. Nodes with high contributions are excellent candidates for external interventions such as drag targets to control the network's state for both PPI networks of \textbf{(h-i)} human and \textbf{(j-k)} yeast.
	}
	\label{fig:cellular_stability}	
\end{figure}

\clearpage
 
 \begin{figure}
 	\centering
 	\includegraphics[width=0.95\textwidth]{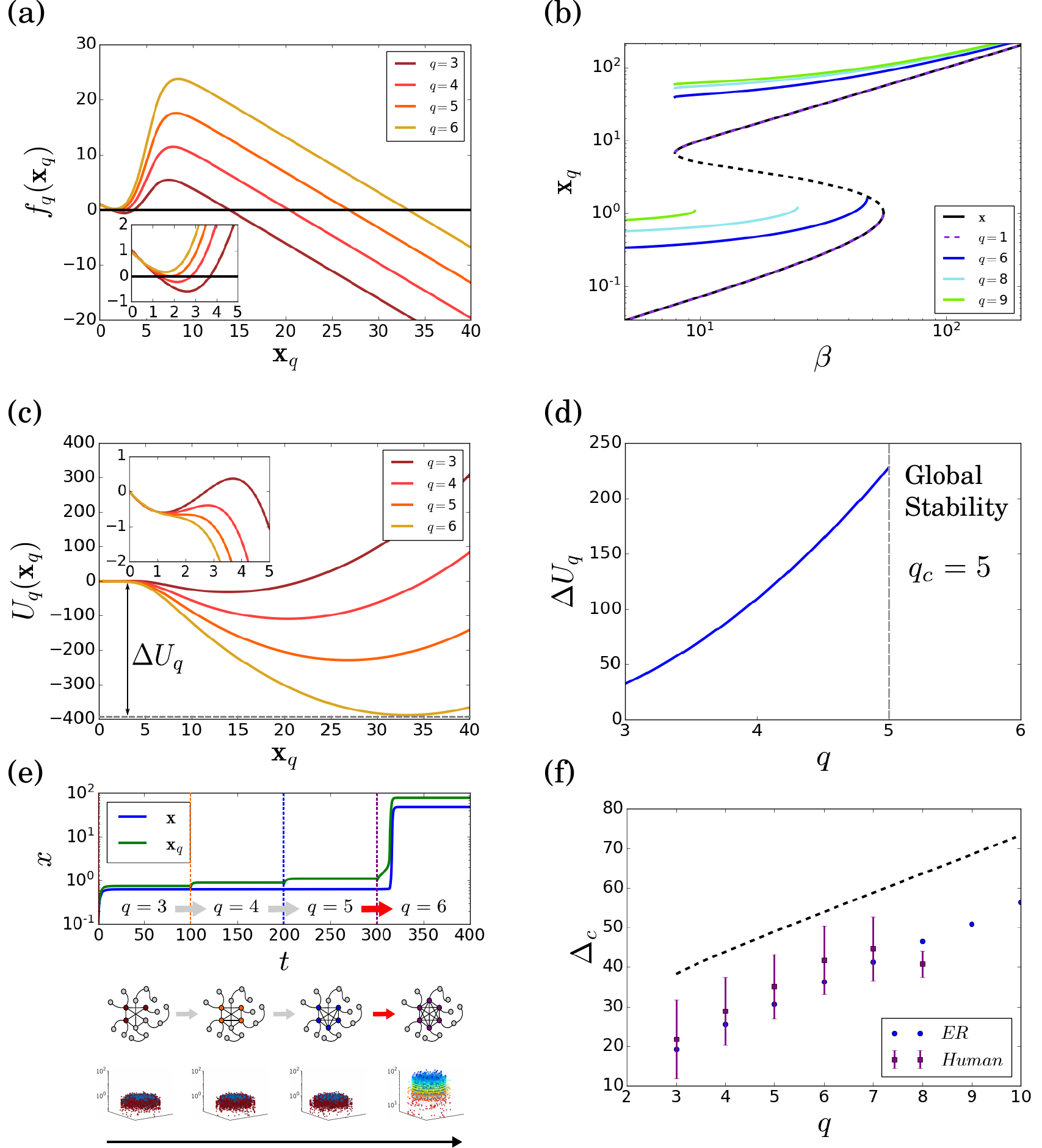}
 	\caption{\footnotesize \textbf{Neuronal dynamics stability (a)} The clique dynamics is described by Eq.~\eqref{eq:neuronal_clique_xq} where stable states are obtained by setting $f_q(\x_q)$ to zero. Inset: zoom in on the region of the low stable state. For large enough cliques, the low stable state disappears. \textbf{(b)} The activity of the stable state increases with the size of the clique and depends on the average weighted degree of the network $\beta$.  \textbf{(c)} The stability of the high stable state can be measured using the clique potential gap $\Delta U_q$ to the nearest maxima, which increases with the clique size. Inset: for large enough cliques, the low stable state disappears and the high stable state becomes globally stable. \textbf{(d)} $\Delta U_q$ increases with the clique size while above a critical value $q_c$ the low stable state disappears, and the high stable state becomes globally stable. \textbf{(e)} When the system is initiated at $\x_0$ and the clique size is below $q_c$ the system will remain at $\x_0$. As the clique's size increases, its activity increases as well. Once its size exceeds $q_c$, the clique spontaneously rises toward $\x_{1,q}$ followed by the entire system. \textbf{(f)} The activity gap $\Delta_c$ which is relevant for clique sizes below $q_c$ increases with clique size and calculated theoretically from .
 	}
 	\label{fig:Neuronal_stability}	
 \end{figure}
 
 \clearpage
 
 \begin{figure}
	\centering
	\includegraphics[width=0.9\textwidth]{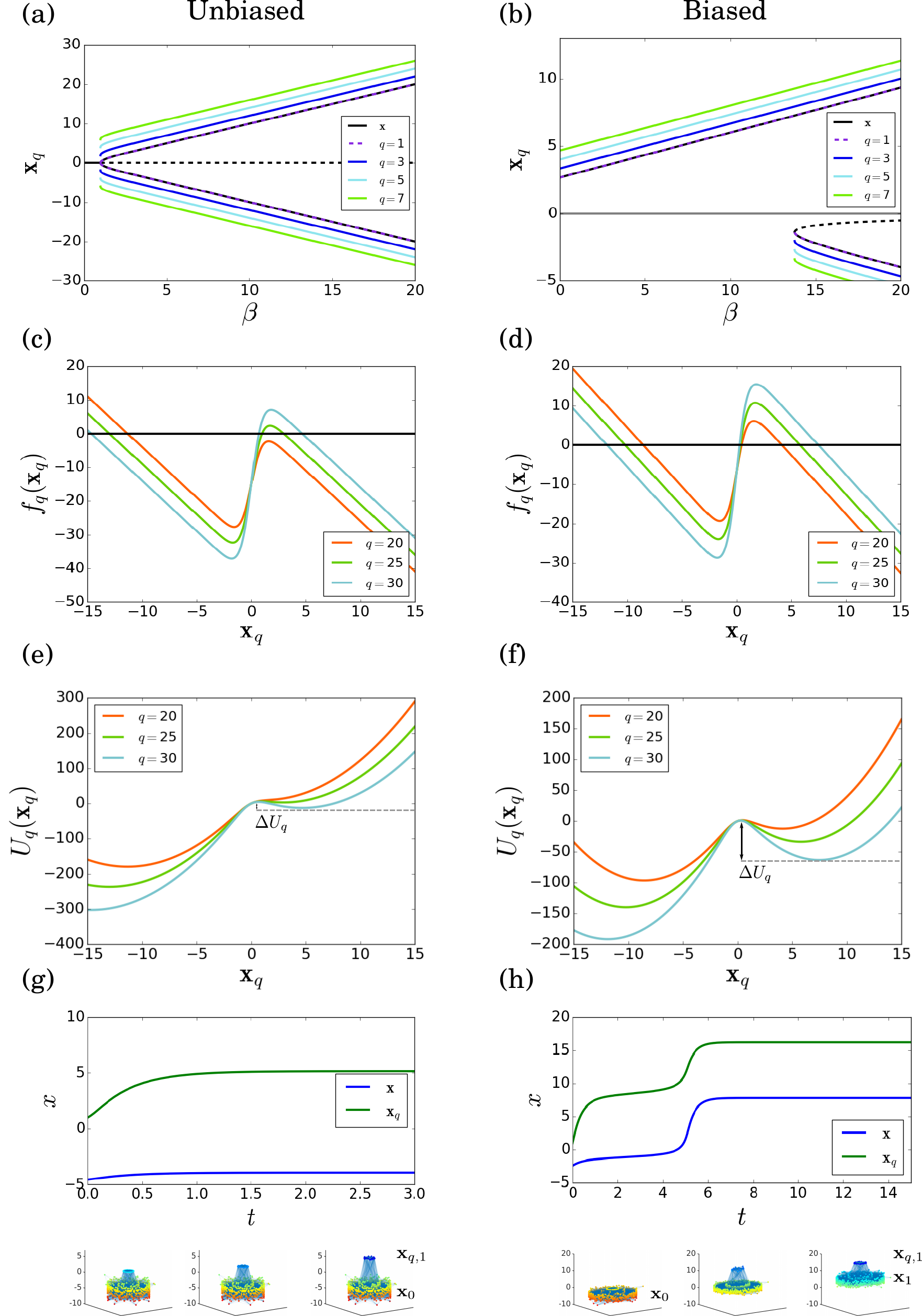}
	\caption{\footnotesize \textbf{Social dynamics stability.} Social dynamics can be \textbf{(a)} unbiased where nodes in both states have the same influence and $|\x_0| = |\x_1|$ and \textbf{(b)} biased where nodes in the higher state have more influence then nodes in the lower state and  $|\x_0| < |\x_1|$. \textbf{(c-d)} The clique dynamics in both cases follow Eq.~\eqref{eq:social_clique_xq} and the stable states are obtained by setting $f_q(\x_q)$ to zero. \textbf{(e-f)} The stability of the states is described by the potential gap $\Delta U_q$ which increases with the clique size. \textbf{(g)} If the clique is initiated in the basin of attraction of $\x_{q,1}$ it will move toward this state, even if the entire system remains at $\x_0$ showing the echo chamber phenomenon. However, in the unbiased case, the cliques are not influential enough to drag the entire system toward $\x_1$. \textbf{(h)} In contrast, in the biased case, nodes around the proffered $\x_1$ state have higher influence. Thus, cliques can spontaneously rise to $\x_{1,q}$ and drag the entire system behind. This case shows how echo chambers can create global opinion change.    }
	\label{fig:Social_stability}	
\end{figure}

\clearpage


{\color{white}. }
\vspace{105mm}
\begin{center}
\textbf{\color{blue} \Large FIGURES - FULL SIZE}
\end{center}

\begin{figure}[b]
\centering
\includegraphics[width=1\textwidth]{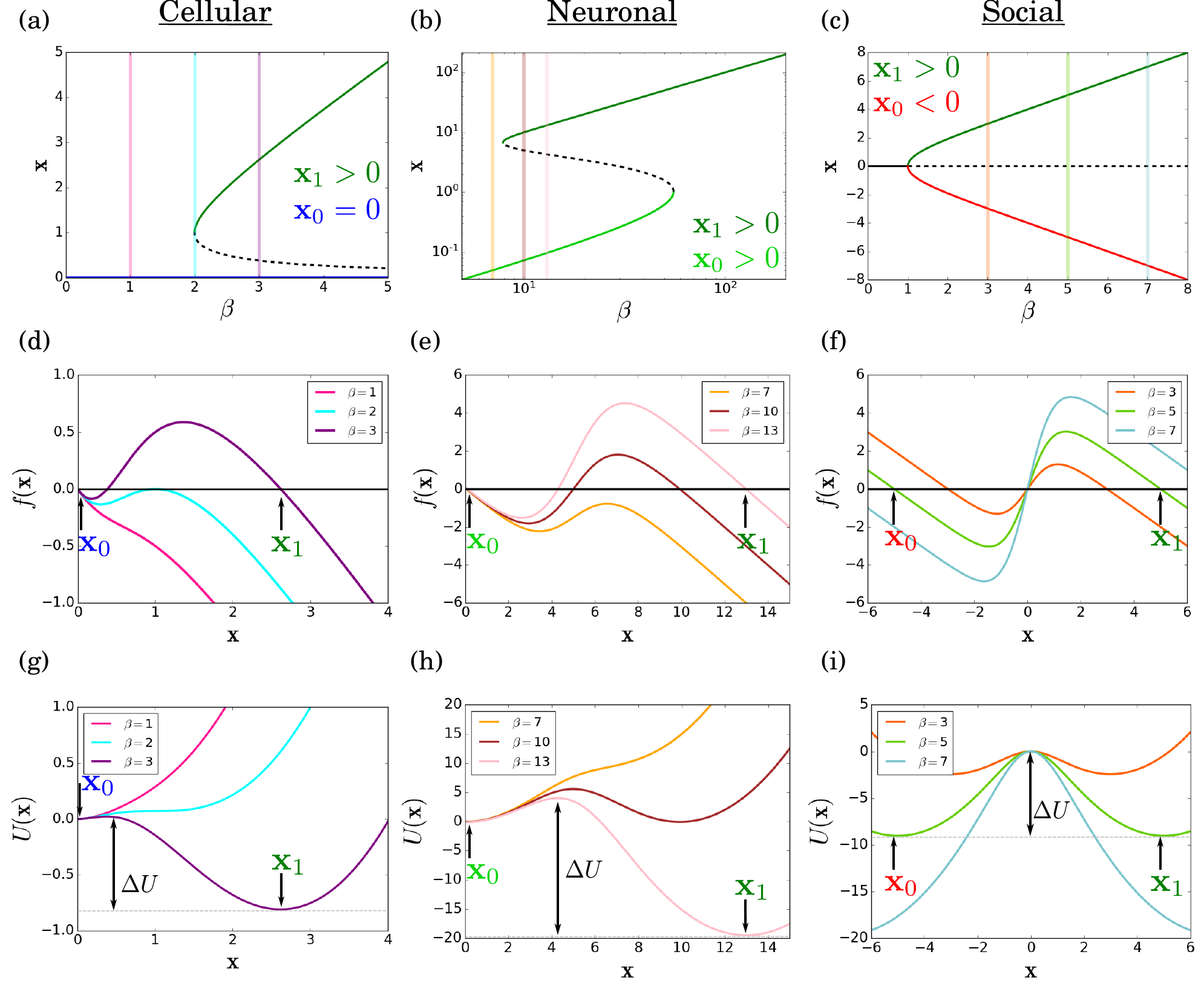}

\textbf{Figure 1.\ Dynamical stability}. 
\end{figure}

\clearpage

\begin{figure}
\centering
\includegraphics[width=1\textwidth]{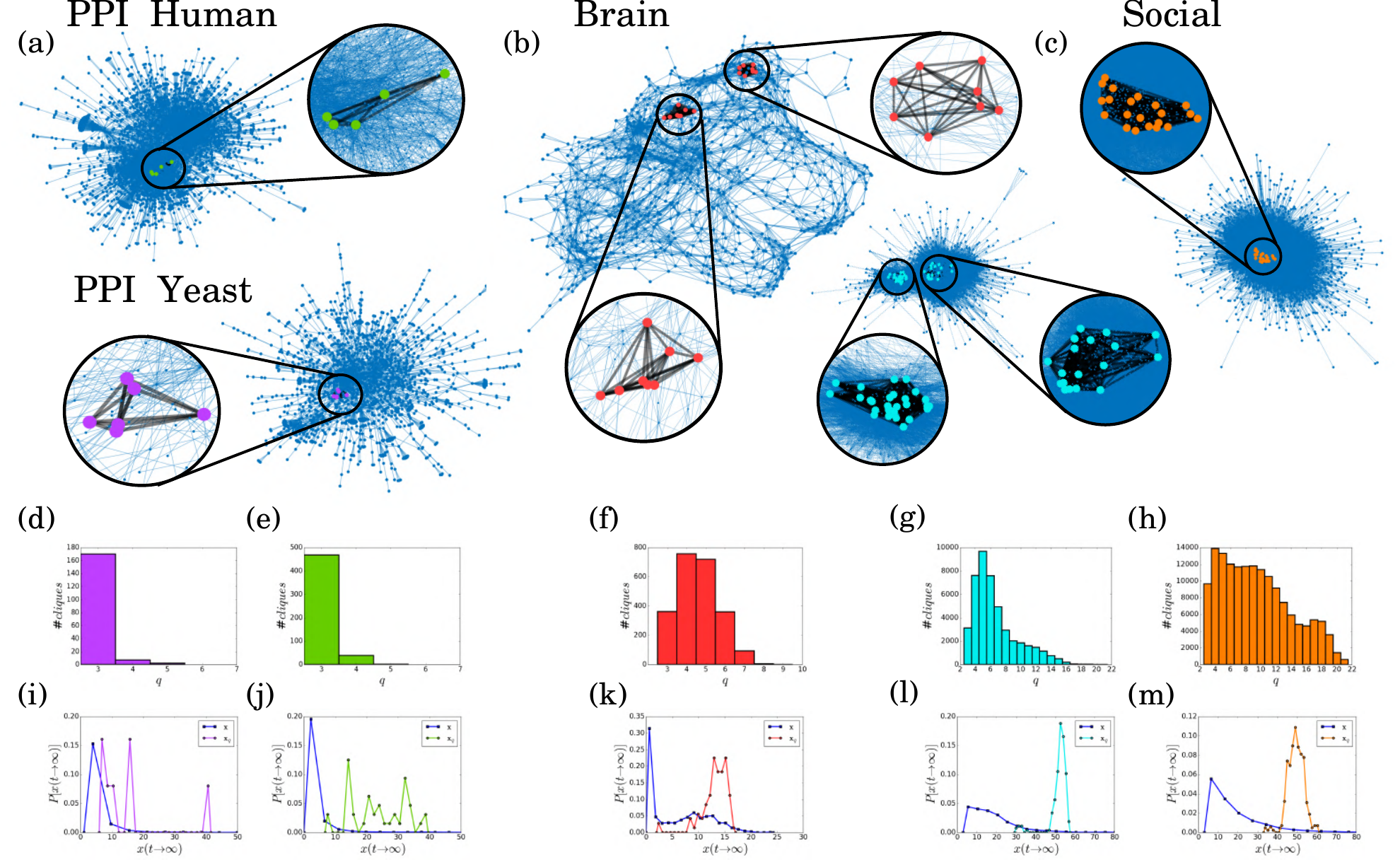}
\textbf{Figure 2.\ Cliques in real-world networks}.
\end{figure}

\clearpage

\begin{figure}
	\centering
	\includegraphics[width=1\textwidth]{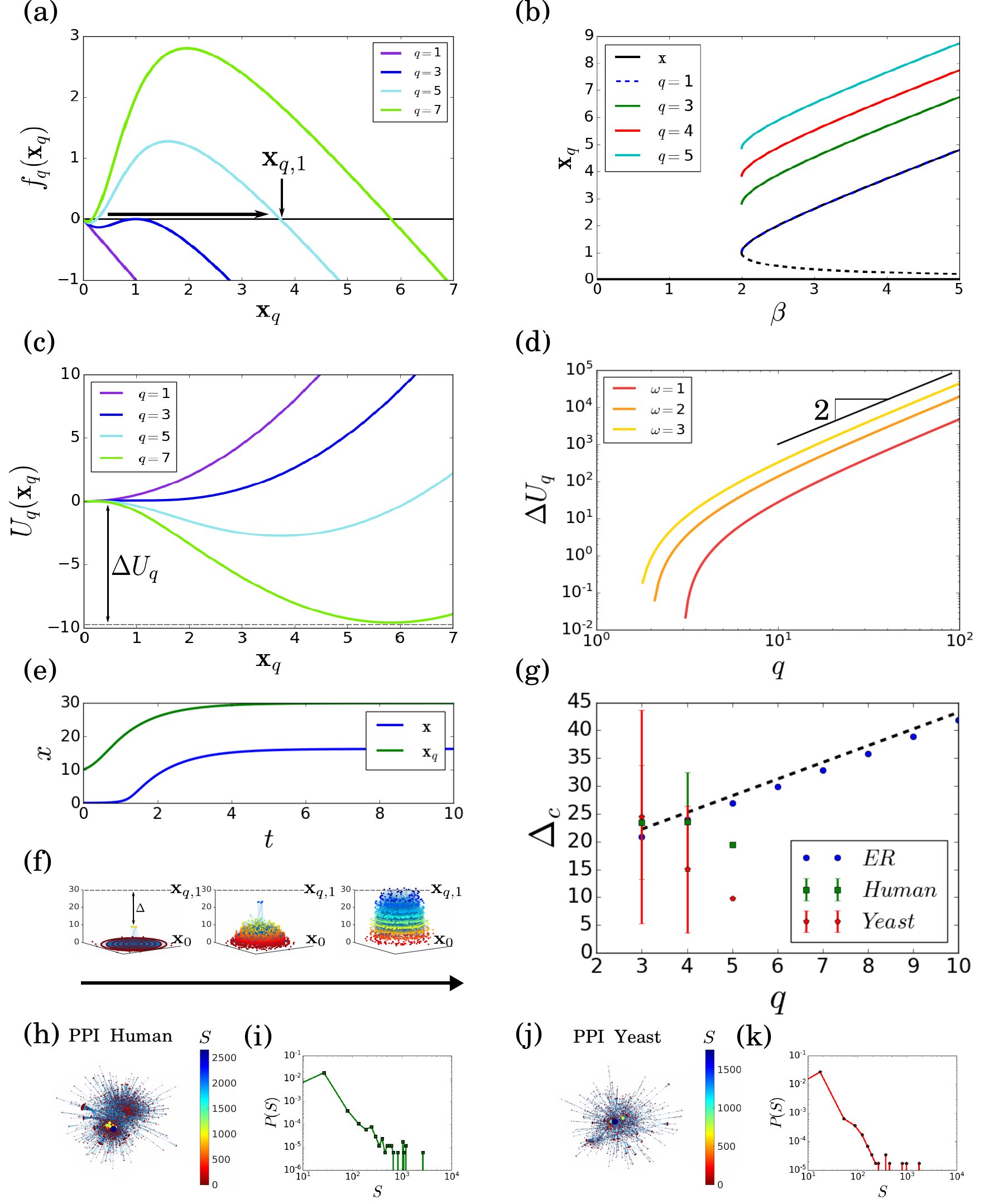}
	\textbf{Figure 3.\ Cellular dynamics stability}.
\end{figure}

\clearpage
 
\begin{figure}
\centering
\includegraphics[width=1\textwidth]{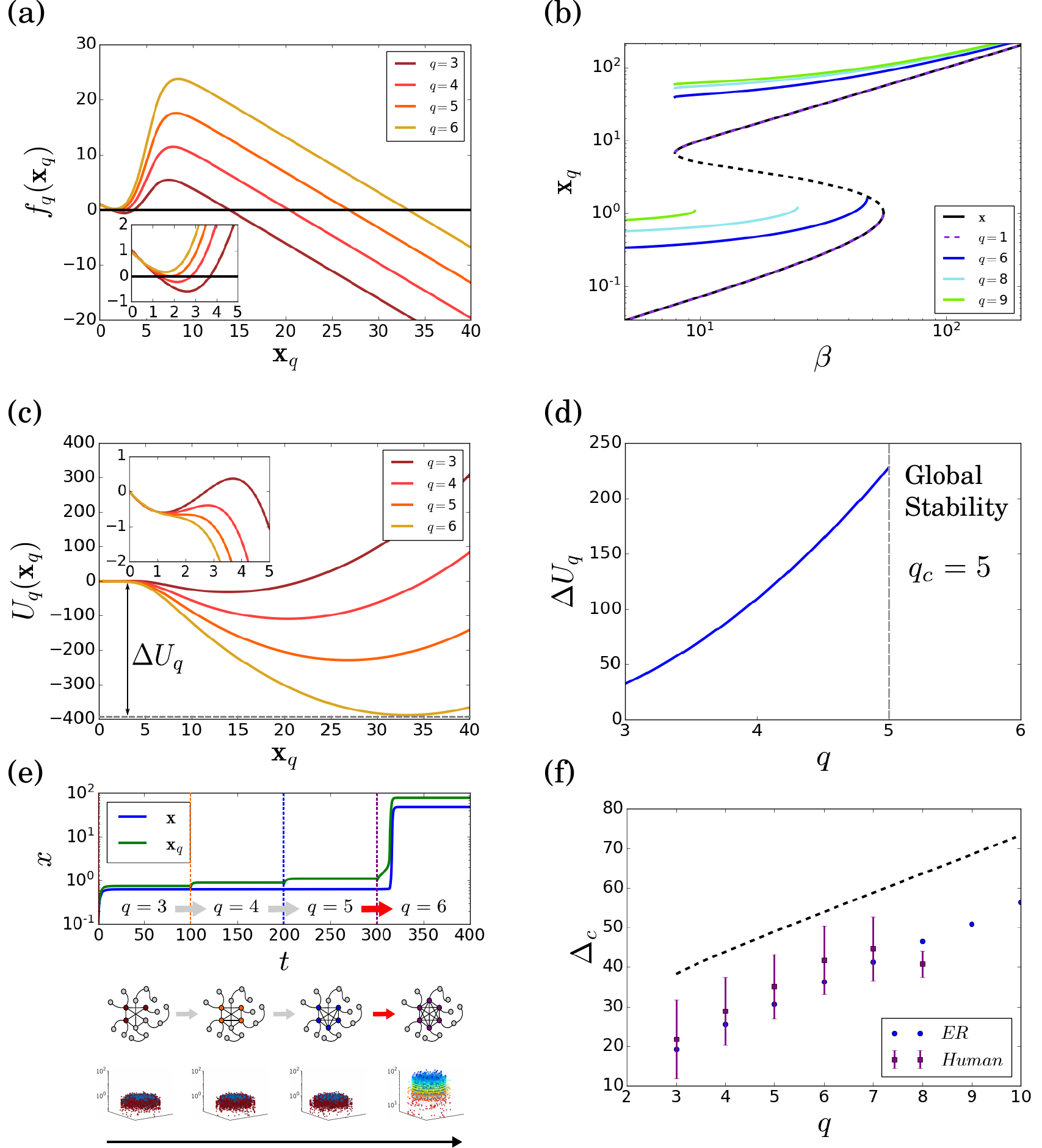}
\textbf{Figure 4.\ Neuronal dynamics stability}.
\end{figure}

\clearpage

\begin{figure}
	\centering
	\includegraphics[width=1\textwidth]{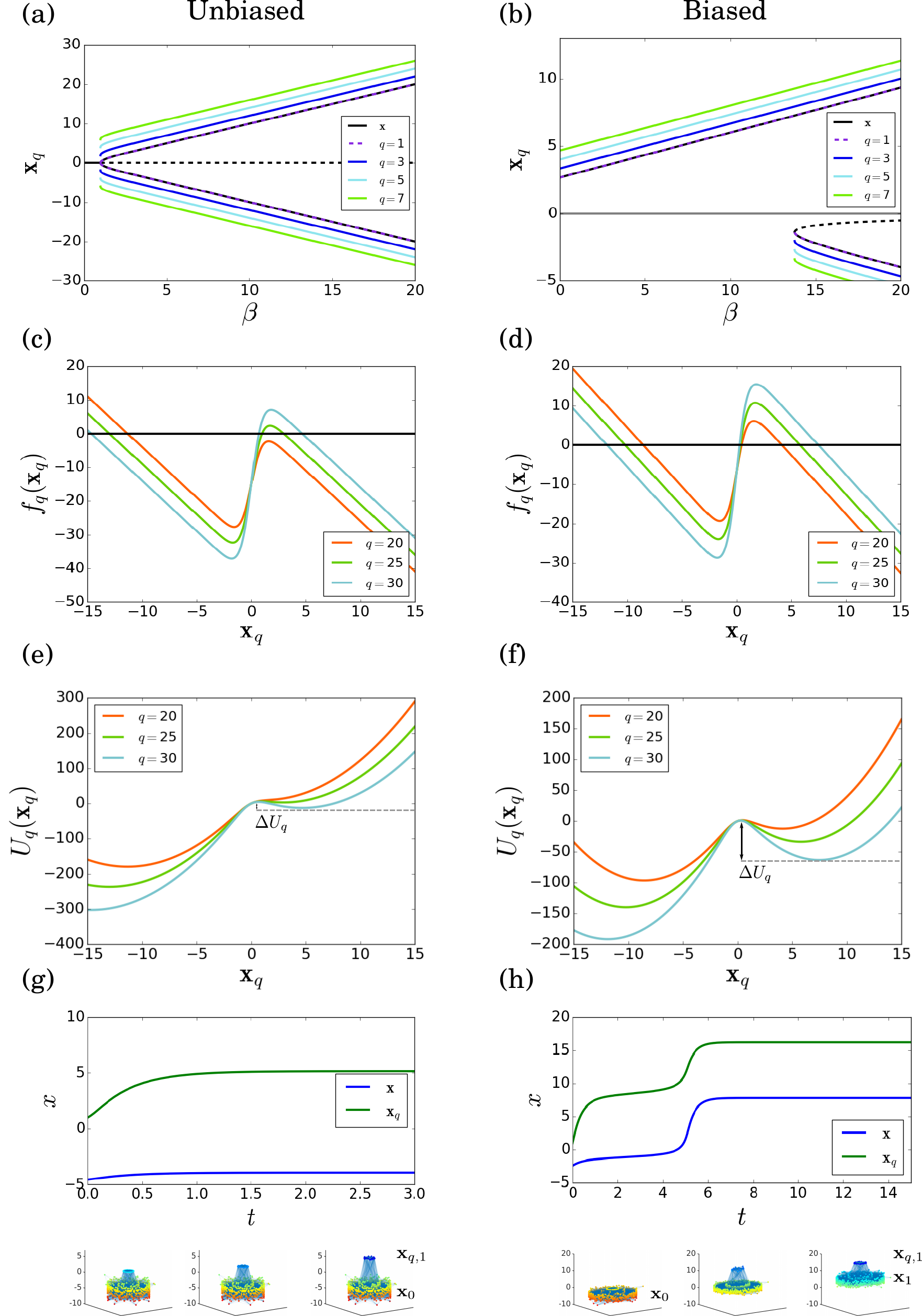}
	\textbf{Figure 5.\ Social dynamics stability}.
\end{figure}

\clearpage

\end{document}